\documentclass{elsart4-1}

\usepackage{amssymb,amsmath}

\usepackage[english,francais]{babel}

\newtheorem{e-proposition}[theorem]{Proposition}

\newtheorem{e-definition}[theorem]{Definition\rm}

\renewcommand{\theequation}{\arabic{equation}}
\def\og{\leavevmode\raise.3ex\hbox{$\scriptscriptstyle\langle\!\langle$~}}
\def\fg{\leavevmode\raise.3ex\hbox{~$\!\scriptscriptstyle\,\rangle\!\rangle$}}

\def\calw{{\mathcal W}}
\def\caln{{\mathcal N}}
\def\calq{{\mathcal Q}}

\def\reals{\mathbb R}
\def\complex{\mathbb C}
\def\zet{\mathbb Z}
\def\CP{\mathbb C\mathbb P}
\def\RP{\reals\mathbb P}

\def\del{\partial}
\def\delbar{\bar\partial}
\def\ii{{\it i}}
\def\ee{{\rm e}}

\def\eg{{\it e.g.}}

\begin{document}

\begin{frontmatter}



\selectlanguage{english}
\title{D-branes from Matrix Factorizations\thanksref{talk}}
\thanks[talk]{Talk given by J.W. at Strings '04, June 28-July 2,Paris}

\vspace{-2.6cm}

\selectlanguage{francais}
\title{}


\selectlanguage{english}
\author[kentaro]{Kentaro Hori}
\ead{hori@physics.utoronto.ca}
\author[johannes]{Johannes Walcher}
\ead{walcher@kitp.ucsb.edu}

\address[kentaro]{University of Toronto, Toronto, Ontario, Canada}
\address[johannes]{Kavli Institute for Theoretical Physics,
University of California, Santa Barbara, California, USA
\thanksref{after}}
\thanks[after]{after September 1st 2004: {\it Institute for Advanced Study,
Princeton, New Jersey, USA}}

\begin{abstract}
B-type D-branes can be obtained from matrix factorizations of
the Landau-Ginzburg superpotential. We here review this promising
approach to learning about the spacetime superpotential of
Calabi-Yau compactifications. We discuss the grading of
the D-branes, and present applications in two examples: the
two-dimensional torus, and the quintic.

\vskip 0.5\baselineskip

\selectlanguage{francais}
\noindent{\bf R\'esum\'e}
\vskip 0.5\baselineskip
\noindent
Les D-branes de type B peuvent \^etre d\'ecrites \`a partir 
de factorisations en matrices du super-potentiel de 
Landau-Ginzburg. On revoit ici cette approche prometteuse 
d'apprendre sur le super-potentiel en espace-temps de 
compactifications de Calabi-Yau. On discute la graduation 
des D-branes, et pr\'esente deux exemples: le tore en deux 
dimensions, ainsi que la quintique.

\vskip 0.5\baselineskip
\noindent{\small{\it Mots-cl\'es~:} D-branes; Factorisations en
matrices; Superpotentiel}
\endkeyword
\end{abstract}
\end{frontmatter}


\selectlanguage{english}

\section{Introduction}

D-branes wrapped on supersymmetric cycles in Calabi-Yau manifolds
have many applications throughout string theory. They have been
studied intensively over the last few years from many different
points of view, and these studies have led to many remarkable
results. One aspect of the problem that is still less understood,
at least in practice, is the (effective 4d spacetime)
superpotential, $\calw$, on the worldvolume of such branes.
$\calw$ being an important quantity for any application,
it is worthwhile to look for new ways of computing it.
(There are other motivations for the kind of investigation
I am undertaking here, but this one should suffice for the
moment.)

In my talk, I want to describe a new approach to studying D-branes
in a certain class of well-known backgrounds, the so-called 
Landau-Ginzburg models \cite{martinec,vawa,vafa1,vafa2}.
Briefly put, this approach amounts to studying the equation
\begin{equation}
Q^2 = W\cdot{\rm id} \,,
\end{equation}
where $W$ is a polynomial, the Landau-Ginzburg superpotential, which
characterizes the closed string background, and $Q$ is an (odd) 
matrix with polynomial entries, describing the open string 
configuration. A solution of this equation is called a matrix 
factorization of $W$.

This approach, the details of which will be discussed momentarily,
was proposed in unpublished form by Maxim Kontsevich, and shown to
correctly describe the relevant physics, in \cite{kali1,bhls,kali2}.
Originally, matrix factorizations go back to \cite{eisenbud},
and they have since then been studied continuously in the context of 
singularity theory, which is in fact the mathematical theory 
underlying Landau-Ginzburg models.

My talk, which is based on \cite{howa}, and some work in progress, 
will focus on certain applications of the formalism in the context 
of $\caln=1$, $d=4$ string compactifications. In particular, I want
to show how matrix factorizations of $W$ can lead to (old and
new) insights about the spacetime superpotential $\calw$, and 
more generally, about the local and global structure of moduli 
spaces of D-branes in Calabi-Yau compactifications.
Other recent work on the subject includes
\cite{kali3,lazaroiu,dia1,hori,dia2,hela,hlw}.

\section{Matrix Factorizations}

\subsection{The Warner problem}

To go back to the origin of the problem, I want to consider an
$\caln=(2,2)$ supersymmetric field theory in two dimensions, of the
type that is the starting point for most of perturbative string
theory. Let me consider in particular the worldsheet superpotential
$W$. To preserve supersymmetry, $W$ is a holomorphic function of
the chiral field variables $\Phi$, which (assuming a flat target
space) are complex functions on $\caln=(2,2)$ superspace with 
coordinates
\begin{equation}
x^+=t+x,x^-=t-x, \qquad
\theta^+,\theta^-,\bar\theta^+,\bar\theta^-
\end{equation}
$\Phi$ satisfies
\begin{equation}
\bar D_\pm \Phi=0\,,
\end{equation}
where 
\begin{equation}
D_\pm = \frac{\del}{\del\theta^\pm} - \ii\bar\theta^\pm\del_\pm
\qquad
\bar D_\pm = -\frac{\del}{\del\bar\theta^\pm} + \ii\theta^\pm\del_\pm
\end{equation}
is the usual covariant derivative. The four supersymmetries are
generated by $\calq_\pm$, $\bar\calq_\pm$, where
\begin{equation}
\calq_\pm = \frac{\del}{\del\theta^\pm} + \ii\bar\theta^\pm\del_\pm
\qquad
\bar\calq_\pm = -\frac{\del}{\del\bar\theta^\pm} -\ii\theta^\pm\del_\pm
\end{equation}

I want to study this field theory on a space with boundary, say a
half-space with boundary at $x^+=x^-=t$. Translational invariance in 
$x$-direction being broken, supersymmetry has to be broken also. In 
geometric terms, this means that superspace acquires a superboundary, 
which is one-dimensional $\caln=2$ superspace, with coordinates
\begin{equation}
t, \theta, \bar\theta \,,
\end{equation}
and identified as superboundary via the equations
\begin{equation}
x^+=x^-=t\,,\qquad \theta^+=\theta^-=\theta\qquad
\bar\theta^+=\bar\theta^- =\bar\theta\,,
\end{equation}
What I have written down here is known as a B-type superboundary,
and is invariant under the B-type supersymmetries
\begin{equation}
\calq = \calq_++\calq_- = \frac{\del}{\del\theta}
+\ii\bar\theta\del_t  \qquad 
\bar\calq=\bar\calq_++\bar\calq_- = -\frac{\del}{\del\bar\theta}
-\ii\theta\del_t
\end{equation}
(The other possible superboundary consistent with supersymmetry and
translational invariance, of A-type, leads to a different problem
with a different solution, and I will not consider it here.)

Now an ordinary local field theory, which is invariant under some 
global bosonic spacetime symmetries, will also be invariant in 
the presence of a boundary under all symmetries that leave the 
boundary invariant. This is no longer true for supersymmetries.
In the case at hand, the F-term
\begin{equation}
\int_\Sigma d^2x d\theta^+d\theta^- W + {\rm c.c}
\end{equation}
exihibits a boundary term under the B-type supersymmetry, which
is the supersymmetry preserved by the boundary
\begin{multline}
\int_\Sigma d^2x d\theta^+d\theta^- \bar\epsilon\bigl(\bar\calq_++
\bar\calq_-\bigr)W =
\text{(using $\bar D_{\pm} W=0$)} \\
=\int d^2x d\theta^+d\theta^- \bigl(-2\ii\bar\epsilon\theta^+
\del_+ -2\ii\bar\epsilon\theta^-\del_-\bigr) W 
= \int_{\del\Sigma}dt d\theta \bigl(-2\ii\bar\epsilon W\bigr)
\end{multline}
The problem associated with these boundary terms under 
supersymmetry variations of superpotentials is known as 
the Warner problem \cite{warner}.

\subsection{Solution of the Warner problem}

There are various ways of dealing with the Warner problem.
(see, \eg, \cite{warner,gjs,hiv,linear,hklm,kali1,bhls}).
One possibility
is to introduce boundary conditions that make the boundary
term vanish. This has some potentially unwanted consequences 
such as spontaneously broken worldsheet supersymmetry. 
Another possibility, which is the topic of present 
interest, is to work with free boundary conditions, and 
to introduce additional degrees of freedom living on 
the boundary, whose supersymmetry variation will cancel 
the boundary term of the bulk variation. The simplest 
possibility is to add a boundary F-term
\begin{equation}
\int_{\del\Sigma} dt d\theta \Gamma(t,\theta) f(\Phi)
\end{equation}
where $f(\Phi)$ is some function of bulk fields (which
is chiral on the boundary), and $\Gamma=\Gamma(t,\theta,\bar
\theta)$ is a fermionic superfield on the boundary which 
fails to be chiral
\begin{equation}
\bar D\Gamma= g(\Phi)|_{\del\Sigma} \,,
\end{equation}
where $g(\Phi)$ is some other holomorphic function of
bulk fields. It is easy to see that the Warner term is 
cancelled if and only if
\begin{equation}
f(\Phi) g(\Phi) = 2\ii W(\Phi)
\end{equation}
This equation, which says that $W$ can be factorized into
$f$ and $g$, is the condition that the boundary superpotential 
preserve $\caln=2$ B-type supersymmetry.

\subsection{Consequences}

Under quantization, the Hilbert space of the boundary fermions 
is simply a two-dimensional vector space $\complex^2$ graded 
by Fermion number. In string theory, one will interpret this 
space as the CP space of a D$\bar{\rm D}$-system, and the 
chiral fields $f(\Phi)$, $g(\Phi)$ as a tachyon configuration 
between the brane and the antibrane. More traditionally, one
can view $f$ and $g$ as some relevant perturbation of a free
boundary condition, much as the bulk superpotential term, $W$. 
Of course, for string theory, one will have to insure that 
the induced boundary RG flow reaches a non-trivial IR fixed 
point, about which I will have more to say a little later.

To find out about the spectrum of open strings, we consider
the system on the strip $x\in[0,\pi]$, with solutions of the 
factorization condition $(f_0,g_0)$ and $(f_\pi,g_\pi)$ as
boundary interactions at the two ends of the strip.
As you might vividly imagine, the supercharge will receive
a contribution from boundary terms, which acts by a graded
commutator
\begin{equation}
\ii \bar Q_{\rm bdy} 
\begin{pmatrix}
a & b \\ c & d
\end{pmatrix}
=
\begin{pmatrix}
0 & f_\pi \\ g_\pi & 0
\end{pmatrix}
\begin{pmatrix}
a& b \\ c & d
\end{pmatrix}
-
\begin{pmatrix}
a & -b \\ -c & d
\end{pmatrix}
\begin{pmatrix}
0 & f_0 \\
g_0 & 0
\end{pmatrix}
\end{equation}
on an open string state with CP structure described by 
$a,b,c,d$. In particular, the spectrum of supersymmetric 
ground states is found by studying the cohomology of the 
operator 
\begin{equation}
\delbar + \ii \bar Q_{\rm bdy}
\end{equation}
acting on $2\times 2$ matrix-valued differential forms,
which for flat target space amounts to studying $2\times 2$
matrices with holomorphic entries in the cohomology of 
$\bar Q_{\rm bdy}$, acting as above.

\subsection{Example}

The simplest example of all this are the minimal models,
with just one variable $\Phi=x$, and superpotential
\begin{equation}
W = x^h
\end{equation}
This superpotential can be factorized as \cite{bhls,kali3}
\begin{equation}
x^h = x^n \cdot x^{h-n} \qquad \text{for $n=0,1,\ldots h$}
\end{equation}
leading to 
\begin{equation}
Q_n = 
\begin{pmatrix} 0 & x^n \\ x^{h-n} & 0 \end{pmatrix}
\end{equation}
We will denote the corresponding boundary condition by $M_n$.
The spectrum of chiral operators between $M_{n_1}$ and $M_{n_2}$ 
consists of even
\begin{equation}
\phi^0_{n_1,n_2,j}(x) =
\begin{pmatrix}
x^{j-\frac{n_1-n_2}{2}} & 0\\ 0 & x^{j+\frac{n_1-n_2}{2}} 
\end{pmatrix}\,,
\end{equation}
and odd operators
\begin{equation}
\phi^1_{n_1,n_2,j}(x) =
\begin{pmatrix}
0 & x^{\frac{n_1+n_2}{2}-j-1} \\ -x^{h-\frac{n_1+n_2}{2}-j-1} & 0
\end{pmatrix}\,,
\end{equation}
where
$$
j=\frac{|n_1-n_2|}{2},\frac{|n_1-n_2|}{2}+1,\ldots,
\min\left\{\frac{n_1+n_2}{2}-1,h+\frac{n_1+n_2}{2}-1\right\}.
$$
As you can see, the branes $n$ and $h-n$ are each others' antibrane,
and the branes $n=0,h$ are trivial, as there are no open strings
between them and any other brane. This spectrum agrees with 
the results derived in the rational conformal field theory 
description of B-type D-branes in $\caln=2$ minimal models.

\subsection{Generalizations}

The construction I have described above can be generalized to
include more that one, let's say $N$, D$\bar{\rm D}$ pairs. As
is well-known, boundary fermions are only available when
$N$ is a power of $2$, but the general case can be described
using the language of superconnections, to which the worldsheet
couples through the super-Wilson line. One can also relax 
the requirement that the target space is topologically 
trivial, and include non-trivial gauge field and tachyon 
configurations \cite{lazaroiu}. In what follows, I will 
continue to assume that the target space is flat $\complex^r$.

\subsection{Summary so far}

Let me now summarize this discussion 
\cite{kali1,bhls,kali2,lazaroiu}. B-type supersymmetry
preserving boundary interactions in a Landau-Ginzburg model with 
polynomial bulk superpotential $W(x_1,\ldots,x_r)$ can be
produced by giving a pair of $N\times N$ matrices 
$f(x_1,\ldots x_r)$, $g(x_1,\ldots,x_r)$ with polynomial 
entries satisfying
\begin{equation}
f\cdot g = g\cdot f = W\cdot{\rm id}_{N\times N}
\label{factorize}
\end{equation}
A solution of this equation is called a matrix factorization
of $W$. The matrices $f$ and $g$ can be thought of as describing 
the tachyon configuration between a stack of $N$ space-filling 
branes and $N$ space-filling antibranes, which annihilate 
everywhere except at the critical points of $W$. In the 
supersymmetry charge, $f$ and $g$ are assembled into the odd 
matrix
\begin{equation}
Q = \begin{pmatrix} 0 & f \\ g & 0 \end{pmatrix}
\end{equation}
in terms of which the requirement of $\caln=2$ worldsheet
supersymmetry can be succinctly written as the equation
\begin{equation}
Q^2 = W\cdot {\rm id}_{2N\times 2N}
\end{equation}
$Q$ acts on open strings by a supercommutator as we have
seen above, and this action squares to zero by virtue of the 
super-Jacobi identity. Open string ground states are found by 
studying the cohomology classes of $Q$ acting on matrices with 
polynomial entries. 
\begin{equation}
\{Q,\Phi\} = 0 \qquad \Phi \equiv \Phi+ \{Q,\Phi'\}
\end{equation}

\section{Matrix Factorizations as D-brane category}

Before discussing more interesting examples, I want to explain
the relevance of such constructions in string theory. 
The general idea is that the set of matrix factorizations provides 
a concrete and particularly simple example of a ``D-brane category'',
where branes are objects, and open strings are morphisms.

The equation $Q^2=W$ is the condition that the boundary coupling
preserve $\caln=2$ supersymmetry. In string theory, this requirement 
is not sufficient for the usual applications, let's say, construction 
of $\caln=1$ supersymmetric compactifications of type II theory 
on a Calabi-Yau with branes and fluxes, which is a starting 
point for many recent discussions in string phenomenology. 

$\caln=2$ worldsheet supersymmetry is sufficient, however, 
for one particularly important part of the story, namely (spacetime)
F-terms. In fact, this follows from the general, so-called 
decoupling statement \cite{bdlr,douglas}, which states 
that of the two topological string theories one can contemplate 
for D-branes wrapped on supersymmetric cycles, the one that can
always be defined once $\caln=2$ supersymmetry is preserved,
controls the worldvolume superpotential. Here this model is
the B-model. The fact that this model controls the F-terms is
known for a long time (BCOV). The other model, here the 
A-model, which only makes sense if a certain familiar 
condition on the $U(1)$ R-charge is satisfied, controls the 
D-flatness conditions on the D-brane worldvolume.

\subsection{The grading}

Let me briefly digress on this aspect of the story. You recall
that for conformal invariance in the bulk, we require the 
Landau-Ginzburg superpotential to be quasihomogeneous:
$$
W(\ee^{\ii\lambda q_i} x_i) = \ee^{2\ii\lambda} W(x_i)\,,
$$
which is equivalent to the existence of a (vector) $U(1)$ 
R-charge, which becomes part of the $\caln=2$ superconformal
algebra in the IR. The equivalent statement on the boundary is
that there must exist an assignement of R-charge on CP spaces
such that $Q$ has charge one
$$
\ee^{\ii\lambda R} Q(\ee^{\ii\lambda q_i} x_i) \ee^{-\ii\lambda R}
= \ee^{\ii\lambda} Q(x_i)
$$
This R-charge provides an additional grading on the space of
open strings,
$$
\ee^{\ii\lambda R}  \Phi(\ee^{\ii\lambda q_i} x_i)
\ee^{-\ii\lambda R} = \ee^{\ii\lambda q_\Phi} \Phi(x_i) 
$$
where $\Phi$ is a matrix with polynomial entries,
and since there is only a finite number of polynomials of 
fixed degree, this condition makes the problem $Q^2=W$ 
effectively finite-dimensional, in contrast to geometric 
versions of the same problem.

The condition that the $U(1)$ charges are integer, which is a
necessary condition for the GSO projection, can be achieved
as usual by orbifolding, which means giving a representation 
of the orbifold group $\Gamma$ on the CP space such the matrix 
factorization is equivariant
\begin{equation}
\gamma \, Q\bigl(\gamma(x_i)\bigr) \gamma^{-1} = Q(x_i)
\end{equation}
(Orbifolding also produces a non-trivial K-theory of the
category of matrix factorizations, which otherwise is
at most torsion.)

Finally, unitarity requires that the R-charges be contained
between $0$ and $\hat c$,
$$
0\le q_\Phi \le \hat c (=3)
$$
which should provide a notion of stability (at the
Landau-Ginzburg point).

One can see from this that matrix factorizations are naturally 
equipped with all the structure for a ``D-brane category'', 
and this finally makes Landau-Ginzburg models a natural place 
to revisit the questions about such categories that have been 
asked many times, in the abstract setting as well as in the 
geometries to which the LG models are connected at large volume. 
(The equivalence with the large volume category has been
discussed in \cite{dia1}.)

\subsection{Deformations}

Here, I want to focus on the spacetime superpotential, 
$\calw$, which one thinks of naively simply as the object
that captures the deformation problem of $Q^2=W$, according
to a point of view that has been taken many times in the
literature, see, \eg, \cite{lazaroiu2,tomasiello,dgjt,kklm}. 
In fact, there are two natural questions in 
this context:

(i) Can we deform $Q$, holding $W$ fixed?

(ii) If we deform $W$, is there a corresponding deformation of $Q$?

There is a traditional answer to (i), which is that the
infinitesimal deformations are given by $H^1$, and the
obstructions by $H^2$ (where the grade is given by the 
R-charge). The usual answer to (ii) is that
there is a map from infinitesimal deformations of closed
strings to obstructions of open strings, and this can also
lead to a superpotential. 

Indeed, writing $Q=Q_0+\varphi\Phi$
and $W=W_0+\psi\Psi$, and assuming $Q_0^2=W_0$, one finds the 
equation
\begin{equation}
Q^2-W= \varphi \{Q,\Phi\} + \varphi^2 \Phi^2 - \psi \Psi
\end{equation}
which expresses what we just said. The point I want 
to emphasize is that it can actually happen that for given bulk
deformation, $\Psi$, this equation has no solution for $\Phi$,
thus providing an example of a brane inducing a potential for
a previously marginal bulk deformation. It should be interesting
to use this kind of mechanism in the context of moduli stabilization.

\subsection{F-terms beyond perturbation theory}

Compared to other approaches, the main advantage of studying 
$\calw$ via the equation $Q^2=W$ is that the problem of 
deformations of solutions, their obstructions, and even the 
global properties of moduli spaces of D-branes is a 
finite-dimensional algebraic problem, at least up to the 
possibility of adding an arbitrary number of brane-antibrane 
pairs. 

The usual approach to a problem of this sort is 
to use perturbation theory, with ansatz
\begin{equation}
Q=Q_0+Q_1+Q_2+\cdots
\end{equation}
where $Q_1=a\in H^1(Q_0)$ and $Q_n$ is of order $n$ with respect 
to $a$. Now solving $Q^2=W$ recursively
\begin{align}
& \{Q_0,Q_2\}+(Q_1)^2=0 \\
& \{Q_0,Q_3\}+\{Q_1,Q_2\}=0\\
& \{Q_0,Q_4\}+\{Q_1,Q_3\}+(Q_2)^2=0 \\
& \qquad\qquad\vdots\\
& \{Q_0,Q_n\}+\sum_{i=1}^{n-1}Q_iQ_{n-i}=0 \\
& \qquad\qquad\vdots
\end{align}
requires gauge fixing at each step. Formally, matrix 
factorizations equip the $\zet$-graded vector space $V^*={\rm 
End}[x_i]$ with a differential $d={\rm ad}Q_0$ with cohomology 
$H^*(Q_0)$. Gauge fixing is a choice of degree $-1$-operator 
$u:V^*\to V^{*-1}$ such that $P:=1-du-ud$ is a projector 
on $H^*$. 

In the unobstructed case, the formal solution is
\begin{align}
& Q_2=-u((Q_1)^2)=-u(a^2) \\
& Q_3=-u\{Q_1,Q_2\}=u\{a,u(a^2)\} \\
& Q_4=-u(\{Q_1,Q_3\}+(Q_2)^2)=-u\{a,u\{a,u(a^2)\}\}
-u(u(a^2)\cdot u(a^2)) \\
&\qquad\qquad\vdots \\
& Q_n=-u\left(\sum_{i=1}^{n-1}Q_iQ_{n-i}\right)
=u\lambda_n(a^{\otimes n}) \\
&\qquad\qquad\vdots
\end{align}
where $\lambda_n:(V^1)^{\otimes n}\to V^2$ is defined
recursively.

\def\m{\mathfrak{m}}
Obstructions are measured by the cohomology classes
\begin{equation}
\m_n([a]^{\otimes n}):=\left[\lambda_n(a^{\otimes n})\right]\in H^2(M_0,M_0).
\end{equation}
($\m_n$ are the higher products of an $A_\infty$ structure.)

In the general (obstructed) case,
\begin{equation}
Q=Q_0+a+\sum_{n=2}^{\infty}u\lambda_n(a^{\otimes n})
\end{equation}
satisfies
\begin{equation}
Q^2=W_0-\sum_{n=2}^{\infty}\m_n(a^{\otimes n})
-\sum_{\ell=1}^{\infty}(-1)^{\ell}
\underbrace{u\Biggl[\Delta Q\,...\,u\Biggl[\Delta Q,}_{\ell}
\sum_{n=2}^{\infty}\m_n(a^{\otimes n})\Biggr]...\Biggr],
\end{equation}
where $\Delta Q=Q-Q_0$. The F-term equations are
(assuming, for simplicity, that $W=W_0$ is fixed)
\begin{equation}
\sum_{n=2}^{\infty}\m_n(a^{\otimes n})=0 \quad\in H^2(Q_0)
\end{equation}

In principle, up to ${\rm dim} (V^1)<\infty$ terms can
appear. In practice, perturbation theory closes on a 
small number of operators, $A_1,\ldots,A_N$. One
can then study the complete problem without resorting
to a perturbative expansion (and without the need for
gauge fixing). Indeed, the ansatz ($n={\rm dim} H^1$) 
\begin{equation}
\Delta Q= Q- Q_0 = \sum_{i=1}^n \phi_i\Phi_i
+\sum_{I=1}^N a_IA_I
\end{equation}
satisfies
\begin{equation}
Q^2-W_0=P(\Delta Q)^2+d(\Delta Q+u(\Delta Q)^2)+udQ^2
\end{equation}
(where we used $P=1-ud + du$). We obtain three equations
\begin{equation}
\label{three}
\begin{split}
 P(\Delta Q)^2&= 0 \\
 d\{\Delta Q+u(\Delta Q)^2\}&= 0 \\
 ud(Q^2) &=0.
\end{split}
\end{equation}
Perturbation theory amounts to neglecting $ud(Q^2)=0$,
which is self-consistent to all orders in the perturbation. 
In general, however, one might obtain new solutions which
are invisible in perturbation theory. Matrix factorizations
are a framework in which the more complete treatment of 
such problems is possible.

\section{Example 1 --- The torus}

A Landau-Ginzburg model for the two-dimensional torus can be built
on the LG potential
\begin{equation}
\bigl( W = x^3+y^3+z^3 +\psi x y z \bigr)/\zet_3
\end{equation}
where $\psi$ is the complex structure parameter of the torus.
Consider the matrix
\begin{equation}
A=
\begin{pmatrix}
\alpha x & \beta z & \gamma y\\
\gamma z & \alpha y & \beta x \\
\beta y & \gamma x & \alpha z
\end{pmatrix}
\label{A}
\end{equation}
We see that
\begin{equation}
\det A = (\alpha^3+\beta^3+\gamma^3) xy z - \alpha\beta\gamma
(x^3+y^3+z^3)
\end{equation}
which is equal to $\lambda W$ with
\begin{equation}
\lambda = -\alpha\beta\gamma
\end{equation}
if and only if
\begin{equation}
\alpha^3+\beta^3+\gamma^3 +\psi \alpha\beta\gamma = 0.
\label{alpha}
\end{equation}
Thus, if we let $B$ be the adjoint of $A$ (the matrix of
subdeterminants) up to a factor,
\begin{equation}
\begin{split}
B &:= \frac 1\lambda \mathop{{\rm adj}}(A) \\
&= -\frac 1{\alpha\beta\gamma}
\begin{pmatrix}
\alpha^2 yz -\beta\gamma x^2 &
\gamma^2 xy-\alpha\beta z^2 &
\beta^2xz-\alpha\gamma y^2 \\
\beta^2xy-\alpha\gamma z^2 &
\alpha^2 yz-\beta\gamma y^2 &
\gamma^2yz-\alpha\beta x^2 \\
\gamma^2 xz -\alpha\beta y^2 &
\beta^2 yz-\alpha\gamma x^2 &
\alpha^2xy-\beta\gamma z^2
\end{pmatrix}
\end{split}
\end{equation}
Then we find \begin{equation}
A B = B A = W \, {\rm id},
\end{equation}
as long as $(\alpha,\beta,\gamma)$ obeys \eqref{alpha} and 
$\alpha\beta\gamma$ is non-zero. The moduli space of this
brane on the torus is thus isomorphic to the torus itself,
as expected of any B-type D-brane on the torus.

What happens as $\lambda\to 0$, where the matrix factorization
becomes naively singular? The trick here is to add a trivial 
brane-antibrane pair
\begin{equation}
f= 
\begin{pmatrix}
-\frac 1\alpha W & 0\\ 0 &A
\end{pmatrix}
\qquad
g=
\begin{pmatrix}
-\alpha & 0 \\ 0 & B
\end{pmatrix}
\end{equation}
and make a gauge transformation on CP spaces that removes the 
singular part of $B$.

\section{Example 2 --- The quintic}

The ``mirror quintic'' Landau-Ginzburg model is
\begin{equation}
\bigl( W = x_1^5 + x_2^5 + x_3^5 + x_4^5 + x_5^5 + \psi x_1x_2x_3x_4x_5
\bigr)/(\zet_5)^4
\end{equation}
At $\psi=0$, an interesting factorization of this $W$ can be obtained 
by taking the tensor product of minimal model factorizations I discussed
before
\begin{equation}
Q_0 = 
\begin{pmatrix}
0 & x_1^2 \\ x_1^3 & 0
\end{pmatrix}
\oplus
\begin{pmatrix}
0 & x_2^2 \\ x_2^3 & 0
\end{pmatrix}
\oplus
\begin{pmatrix}
0 & x_3^2 \\ x_3^3 & 0
\end{pmatrix}
\oplus
\begin{pmatrix}
0 & x_4^2 \\ x_4^3 & 0
\end{pmatrix}
\oplus
\begin{pmatrix}
0 & x_5^2 \\ x_5^3 & 0
\end{pmatrix}
\end{equation}
It turns out that this has exactly one marginal operator
\begin{equation}
\Phi = 
\begin{pmatrix}
0 & 1 \\ -x_1 & 0
\end{pmatrix}
\otimes
\begin{pmatrix}
0 & 1 \\ -x_2 & 0
\end{pmatrix}
\otimes
\begin{pmatrix}
0 & 1 \\ -x_3 & 0
\end{pmatrix}
\otimes
\begin{pmatrix}
0 & 1 \\ -x_4 & 0
\end{pmatrix}
\otimes
\begin{pmatrix}
0 & 1 \\ -x_5 & 0
\end{pmatrix}
\end{equation}
as well as one obstruction
\begin{equation}
\Psi = x_1x_2x_3x_4x_5 \cdot {\rm id}
\end{equation}
which is exactly the marginal bulk deformation.
One can also easily see that
\begin{equation}
\Phi^2 = -\Psi
\end{equation}
So, $Q=Q_0+\varphi\Phi$ will square to $W=W_0+\psi\Psi$ iff
\begin{equation}
\varphi^2 + \psi = 0
\end{equation}
This is the F-flatness equation on the D-brane worldvolume, 
which one may integrate to the superpotential
\begin{equation}
\calw = \frac13 \varphi^3 + \varphi\psi
\end{equation}
(This confirms a prediction of \cite{bdlr,brsc}.)
Treating $\psi$ as a closed string parameter, we learn that 
except at $\psi=0$, our brane has two supersymmetric vacua. 
The coalescence of the two vacua at $\psi=0$ is accompagnied 
by the appearance of an additional massless open string field 
$\Phi$.

\subsection{A mirror symmetry interpretation}

In conclusion, I want to offer a geometric interpretation
involving the mirror geometry, which here is the Fermat 
quintic in $\CP^4$
\begin{equation}
X = \{z_1^5+z_2^5+z_3^5+z_4^5+z_5^5=0\}\subset \CP^4
\label{quintic}
\end{equation}
Closed string mirror symmetry gives the map between the marginal
closed string operator $\Psi$ and the generator of $H^2(X,\zet)$,
or in other words, between the complex structure parameter $\psi$
and the K\"ahler parameter $t$ of $X$. The Yukawa coupling which
on the B-model side can be computed to be
\begin{equation}
\kappa_{\psi\psi\psi} = \frac{1}{5^5 + \psi^5}
\end{equation}
can then be expanded on the A-model side
\begin{equation}
\kappa_{ttt} = 5 + \sum \text{(worldsheet instantons)}
\end{equation}
where the $5$ is the result from classical geometry, and
the sum over holomorphic spheres in $X$ is a power
series in $q=\exp(-t)$.

What about open string mirror symmetry? One can show 
\cite{bdlr,bhhw} that the brane I have been 
discussing in the Landau-Ginzburg model is mirror to a 
familiar special Lagrangian cycle in $X$, namely the real 
locus of $X$,
\begin{equation}
L =\{z\in X,z_i\in\reals \;\forall i\}
\label{slag}
\end{equation}
$L$ is topologically an $\RP^3$. Since $H_1(\RP^3)=\zet_2$,
a brane wrapped on $\RP^3$ has two vacua, distinguished by
a discrete Wilson line. These can be identified with the
two vacua we had found for the Landau-Ginzburg brane.
To identify the field that becomes massless at $\psi=0$,
we have to recall that the zero modes of strings wrapped
on special Lagrangians is determined not by ordinary
cohomology but by Floer cohomology, which differs from 
ordinary cohomology by a sum over holomorphic discs ending
on $L$. In our context, the relevant complex is
\begin{equation}
C^0 \overset{0}\longrightarrow C^1
\overset{\delta}
\longrightarrow C^2\overset{0}\longrightarrow C^3
\end{equation}
where for ordinary cohomology, all $C^i\cong\zet$,
and $\delta=2$. For Floer cohomology, we will have
(after tensoring the complex with the appropriate
coefficient field)
\begin{equation}
\delta = 2 + \sum (\text{holomorphic discs})
\end{equation}
where the sum over holomorphic discs should be a power
series in $q^{1/2}=\exp(-t/2)$, because the discs have
half the volume of the spheres. At large volume
$t\rightarrow\infty$, the cohomology is trivial except 
in degree $0$ and $3$. But if $\delta=0$ for some value 
of $t$ (after analytic continuation), there will be an 
additional massless field in degree $1$. The conjecture 
is that this is precisely what happens at $\psi=0$, the 
additional massless field being $\Phi$.

Of course, making this proposal more precise depends on 
identifying the correct map between the (generally) massive 
field $\Phi$ on the Landau-Ginzburg side and the integral 
generator of $C^1$ in the Floer complex.

\section{Summary and Outlook}

We have seen that matrix factorizations are a useful 
B-type model for topological D-branes on Calabi-Yau 
manifolds. In particular, they allow simple calculations
of worldvolume superpotentials for D-branes wrapped
on supersymmetric cycles. When combined with orientifolds
(see, \eg, \cite{bhhw}), this will allow a rather more 
systematic investigation of some properties of $\caln=1$ 
string vacua. We have also seen that the approach
holds some promises toward realizing open string
mirror symmetry for compact Calabi-Yau manifolds
(see \cite{bhlw} for some recent progress in this
direction).

The computation of the superpotential also
illustrates that one may view $Q^2=W$ as a 
finite-dimensional model of (background-independent, 
topological) string field theory, which would be
interesting to explore further. 

One aspect of the story that we have not mentioned
here is the direct connection to geometry in the
B-model. There exist mathematical constructions 
that relate matrix factorizations of $W$ to bundles 
on the hypersurface $\{W=0\}$ in the corresponding 
weighted projective space. It would be interesting to
realize such connections via physical models of the 
linear sigma model type.





\section*{Acknowledgements}
J.W.\ would like to thank C.~Bachas, E.~Cremmer, and P.~Windey 
for the invitation to speak at Strings '04, a most wonderfully 
organized conference.
The work of K.H.\ was supported by Alfred P. Sloan Foundation,
Connaught Foundation and NSERC. 
The research of J.W.\ was supported in part by the National Science 
Foundation under Grant No.\ PHY99-07949.

\end{document}